\documentclass[aps,prl,groupedaddress,twocolumn,showpacs]{revtex4}
\usepackage{graphicx,graphics,psfrag,amsmath,calc,amssymb}

\begin{document}
\title{Time evolution and matter wave interference in Fermi condensates}
\author{Wei Zhang}
\author{C. A. R. S\'{a} de Melo}
\affiliation{School of Physics, Georgia Institute of Technology,
Atlanta, Georgia 30332}

\date{\today}

\begin{abstract}
We discuss matter wave interference of Fermi condensates in strongly coupled $s$-wave and $p$-wave channels
and the time evolution of a single cloud upon release from trap.
In $s$-wave systems, where the order parameter is a complex scalar, we find that the interference patterns
depend on the relative phase of the order parameters of the condensates. In $p$-wave systems
involving the mixture of two-hyperfine states,
we show that the interference pattern exhibits a polarization effect depending
on the relative orientation of the two vector order parameters.
However, in $p$-wave systems involving a single hyperfine state, we show that
this angular effect reduces to an overall phase difference between
the two interfering clouds, similar to $s$-wave.
Lastly, we also point out that $p$-wave Fermi condensates exhibit
an anisotropic expansion, reflecting the spatial anisotropy of the underlying
interaction between fermions and the orbital nature of the vector order parameter.

\end{abstract}

\maketitle

Matter-wave interference is a very powerful tool to study quantum
phase coherence between atomic Bose Einstein condensates
(BEC)~\cite{shin-04,schumm-05}, and spatial quantum noise of
bosons in optical lattices~\cite{bloch-05}. Similar techniques can
also be applied to study Fermi
condensates,~\cite{hulet-03,greiner-03,zwierlein-03,bartenstein-04,bourdel-04,thomas-04}
where superfluidity can be tuned from the BCS to the BEC regime.
% as a function of magnetic field.
These experiments may reveal that the time dynamics in the BCS regime is overdamped (large Cooper
pairs can decay into two atoms), while in the BEC regime it is essentially undamped
(tightly bound molecules are stable)~\cite{sademelo-93,iskin-06}.
Matter-wave interference experiments of $s$-wave Fermi condensates may be readily performed,
since stable condensates already exist. For $s$-wave Fermi condensates in the BEC regime quantum
interference effects are expected to be similar to those of atomic Bose condensates,
and the interference pattern should depend essentially on the phase difference of the order parameters between
two interfering clouds.

In contrast, it is more interesting to study interference effects in
$p$-wave superfluids because of the vector nature of the order parameter.
Many groups have reported some progress towards the formation of
$p$-wave Fermi condensates in single clouds~\cite{regal-03b,zhang-04,schunck-05}
and in optical lattices~\cite{gunter-05}, where $p$-wave Feshbach resonances have been observed.
For Feshbach resonances currently tried in single clouds atom losses have been significant,
and the realization of stable $p$-wave condensates has not been achieved yet.
However, other unexplored Feshbach resonances in single clouds may
show less dramatic two-body dipolar or three-body losses as observed in optical lattices~\cite{gunter-05}.
Thus, these experimental difficulties may be surpassed in the immediate future.

We discuss in this manuscript time dynamics and matter-wave interference
of $s$-wave and $p$-wave Fermi condensates.
Our main results are as follows. While in atomic BEC and $s$-wave Fermi superfluids quantum interference
patterns depend essentially on the relative phase of the two clouds,
we find that in the $p$-wave Fermi superfluids there can also be a strong dependence on the relative angle
between the two vector order parameters, thus producing a polarization effect.
Furthermore, we show that $p$-wave Fermi condensates exhibit
an anisotropic expansion, reflecting the spatial anisotropy of the underlying
interaction between fermions and the orbital nature of the vector order parameter.

We consider a system of fermions with mass $m$ in two hyperfine
states (pseudospins), labeled by greek indices $\alpha = 1,2$. The
Hamiltonian density is (with $\hbar = k_B =1$)
\begin{eqnarray}
\label{eqn:hamiltonian}
&&H ({\bf r}, t) = \psi_{\alpha}^\dagger({\bf r},t)
\left[-\frac{\nabla_{\bf r}^2}{2m} + U_{\rm ext}({\bf r}, t) \right]
\psi_{\alpha}({\bf r},t)
\nonumber \\
&&
- \int d{\bf r}' [ \psi_\alpha^\dagger({\bf r},t)
\psi_\beta^\dagger({\bf r}',t)V_{\alpha\beta\gamma\delta}({\bf r}-{\bf r}')
\nonumber
\psi_\gamma({\bf r}',t) \psi_\delta({\bf r},t)],
\nonumber
\end{eqnarray}
where repeated greek indices indicates summation,
$\psi_\alpha^\dagger$ ($\psi_\alpha$) are creation (annihilation)
operators of fermions in state $\alpha$, and $H (t) = \int d {\bf
r} H ({\bf r}, t)$ is the Hamiltonian. The trapping potential is
$U_{\rm ext}({\bf r},t) = \sum_{j=x,y,z} \omega_j(t) r_j^2/2$,
where $\omega_j(t<0) = \omega_{j,0}$ are constants.

The generating functional for non-equilibrium processes associated with $H(t)$ is~\cite{tokatly-04}
\begin{equation}
\label{eqn:partition-general}
Z(t) = {\rm Tr}  \hat{\cal U}^\dagger (t, t_0)
\exp\left[-\beta(H(t_0)- \mu_\alpha N_\alpha)\right]
\hat{\cal U}(t,t_0),
\nonumber
\end{equation}
where $N_\alpha$ is the number operator for fermions of type $\alpha$,
$\mu_\alpha$ is the corresponding chemical potential,
and $\beta = 1/T$ is the inverse temperature.
Here, $\hat{\cal U}(t,t_0) \equiv \exp[-i\int_{t_0}^t H(t')dt']$
is the time evolution operator.
This expression implicitly implies that the system is in thermal equilibrium
at any time $t_0<0$, since
$Z(t_0) = {\rm Tr} \exp\left[-\beta(H(t_0)- \mu_\alpha N_\alpha)\right]$.
By introducing a complex time $\tau$, the generating functional can be written as
\begin{equation}
\label{eqn:partition}
Z(t) = \int_{\rm BC} \Pi_\alpha D[\psi_\alpha^\dagger, \psi_\alpha]
e^{-i [S_2 (\psi_j^\dagger, \psi_j) + S_4(\psi_j^\dagger, \psi_j)]},
\end{equation}
where the boundary condition (BC) of the functional integral
is $\psi_\alpha({\bf r}, t_0 - i/T ) = -\psi_\alpha({\bf r}, t_0)$.
The quadratic term is
\begin{equation}
\label{eqn:gaussian}
S_2(\psi_j^\dagger, \psi_j)
=
\int_C d\tau \int d{\bf r} \psi_\alpha^\dagger ({\bf r},\tau)
\hat{\cal L}_\alpha ({\bf r}, \tau) \psi_\alpha ({\bf r},\tau),
\end{equation}
where the integration contour $C$ is shown in Fig. 1.
The one-particle operator ${\cal L}_\alpha$ is defined as
$$
\hat{\cal L}_\alpha ({\bf r}, \tau) = -i \partial_\tau - \frac{\nabla^2}{2m}
+ U_{\rm ext}({\bf r},\tau) -\mu_\alpha.
$$
\begin{figure}
\begin{center}
\psfrag{Retau}{Re($\tau$)}
\psfrag{Imtau}{Im($\tau$)}
\psfrag{-b}{$-\beta$}
\psfrag{t}{$t$}
\psfrag{t0}{$t_0$}
\psfrag{C}{$C$}
\psfrag{0}{$0$}
\includegraphics[width=5.0cm]{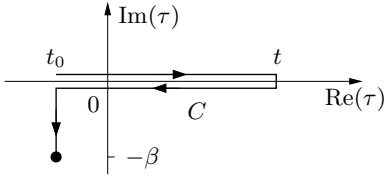}
\end{center}
\caption{Integration contour $C$ used in Eq. (\ref{eqn:gaussian}).}
\label{fig:phase}
\end{figure}

In what follows, we discuss the $p$-wave case in detail and quote
the more standard results for $s$-wave. To calculate the fourth
order term $S_4$, we write $V_{\alpha\beta\gamma\delta}({\bf x}) =
V({\bf x}) \Gamma_{\alpha\beta\gamma\delta}$, where in a triplet
channel $\Gamma_{\alpha\beta\gamma\delta}= {\bf v}_{\alpha\beta}
\cdot ({\bf v}^\dagger)_{\gamma\delta}$. The pseudospin matrices $
(v_j)_{\alpha\beta} \equiv (i \sigma_j \sigma_y)_{\alpha\beta}$,
where $\sigma_j$ are Pauli matrices. This form of interaction
implies that the interaction strength is the same for all
pseudospin channels. (Deviations from this assumption will be
discussed later). By introducing ${\bf B}^\dagger({\bf r}, {\bf
r}' ,\tau) = \psi_\alpha^\dagger({\bf r},\tau) {\bf
v}_{\alpha\beta} \psi_\beta^\dagger({\bf r}',\tau)$ and the
corresponding auxiliary field ${\bf d}^\dagger({\bf r}, {\bf r}',
\tau)$, the interaction term $S_4$ can be decoupled via a
Hubbard--Stratonovich transformation. We integrate out the
fermions and write ${\bf d}({\bf r}, {\bf r}', \tau) = \sum_n {\bf
D}_n({\bf R},\tau) \eta_n({\bf x})$, where $\eta_n({\bf x})$ are
eigenfunctions of the reduced two-body Hamiltonian ${\cal H}_2 =
-\nabla_{\bf x}^2/m + V({\bf x})$. Here, ${\bf R}=({\bf r}+{\bf
r}')/2$ and ${\bf x}={\bf r}-{\bf r}'$ are center-of-mass and
relative coordinates, respectively. In strong coupling limit where
the interaction is the largest energy scale in the problem (with
typical strength much larger than temperature and trapping
potential), the ground state of ${\cal H}_2$ has the same symmetry
of $V$. For definiteness, we consider a pure $p$-wave interaction
where the ground state is three-fold degenerate and labeled by
$\nu=x,y,z$, with corresponding eigenfunctions $\eta_0(x)
\hat{x}_\nu$. Thus, at low temperatures where the higher energy
states are not excited, we have ${\bf d} = \sum_{\nu} {\bf
D}_{\nu} ({\bf R},\tau) \eta_{0}(x) \hat{x}_\nu$.

In this strong coupling limit, Cooper pairs are tightly bound molecules
and their relative degrees of freedom can be integrated out leading to
\begin{eqnarray}
\label{eqn:effaction}
S_{\rm eff} &=& - \int_C d\tau \int d{\bf R}
\Big \{
{\bf D}^\dagger({\bf R},\tau) \cdot \left[\hat{K}{\bf D} ({\bf R},\tau)\right]
\nonumber \\
&& - \frac{g_0}{2} \left[ 2 \vert {\bf D} ({\bf R},\tau)\vert^4 -
\vert {\bf D}^2({\bf R},\tau)\vert^2 \right] \Big\},
\end{eqnarray}
where $D_j ({\bf R}) \equiv \sum_{\nu} D_{j,\nu}({\bf R})$
is the order parameter,
and the operator $\hat{K} = i\partial_\tau - 2 U_{\rm ext}({\bf R},\tau)
+ \nabla_{\bf R}^2 / (4m)$ corresponds to the action of an ideal
non-equilibrium gas of Bose particles with mass $M=2m$. This action leads to equations of motion
\begin{eqnarray}
\label{eqn:GP}
i\partial_t D_j &=& \left[ -\frac{\nabla_{\bf R}^2}{2M} + 2 U_{\rm ext} ({\bf R},t)
+ 2 g_0 \vert {\bf D} \vert^2 \right] D_j
\nonumber \\
&&
\hspace{1cm}
- g_0 \left({\bf D}\cdot {\bf D} \right) D_j^\dagger.
\end{eqnarray}
Notice that this expression is different from the time-dependent
Gross-Pitaevskii (TDGP) equation for a vector boson field. The
difference comes from the last term, which describes a non-unitary
complex order parameter of the underlying paired fermions. In
contrast, the standard TDGP equation for scalar bosons is obtained
in the $s$-wave case~\cite{tokatly-04}.

Equation (\ref{eqn:GP}) can be simplified to the TDGP form in two
special cases. First, if two hyperfine states are equally
populated and ${\bf D}$ is unitary, then it can be represented by
a real vector with an overall phase factor, leading to the
equation of motion
\begin{equation}
\label{eqn:GPunitary}
i\partial_t D_j = \left[ -\frac{\nabla_{\bf R}^2}{2M} + 2 U_{\rm ext} ({\bf R},t)
+ g_0 \vert {\bf D} \vert^2 \right] D_j.
\end{equation}
Second, if only one hyperfine state is populated, then ${\bf D}$ is non-unitary
and ${\bf D} = A (1, -i, 0)$, where $A$ is a complex constant.
Thus, the last term ${\bf D}\cdot {\bf D}$ in Eq.~(\ref{eqn:GP}) vanishes,
and the equation of motion is identical to Eq.~(\ref{eqn:GPunitary}),
with $g_0 \to 2g_0$. In the following discussion, we confine ourselves
to these two special cases.

For definiteness, we assume that the trapped Fermi superfluid is released at time $t=0$,
i.e., $U_{\rm ext}(t>0)=0$. Thus, for $t<0$, the system is described by
\begin{equation}
\label{eqn:staticGP}
\mu_0 D_j({\bf R}) =
\left[ -\frac{\nabla_{\bf R}^2}{2M} + 2 U_{\rm ext} ({\bf R})
+ g \vert {\bf D} \vert^2 \right] D_j,
\end{equation}
where $\mu_0$ is the effective boson chemical potential, $g=g_0$
for the unitary case corresponding to two hyperfine states, and
$g=2g_0$ for a non-unitary case corresponding to a single
hyperfine state. When the composite Boson interactions are
dominant the Thomas--Fermi (TF) approximation leads to $\vert {\bf
D}({\bf R},0) \vert = g^{-1/2}[\mu_0 - 2U_{\rm ext}({\bf
R})]^{1/2}$ for $\mu_0 \ge 2U_{ext}({\bf R})$, and $\vert {\bf
D}({\bf R},0) \vert = 0$, otherwise. When this approximation fails
the initial condition for the time evolution can be obtained by
solving Eq.~$(\ref{eqn:staticGP})$ numerically.

For $t >0$, we use the transformation $ \label{eqn:scale}R_j(t) =
b_j(t) R_j(0)$, where $b_j(t)$ are scaling factors
satisfying,~\cite{kagan-96,castin-96}
\begin{equation}
\label{eqn:bj} \frac{d^2 b_j(t)}{dt^2} =
\frac{\omega_{j,0}^2}{A(t) b_j(t)}
\end{equation}
with initial conditions $b_j(t<0) \equiv 1$ for all $j$,
and $A(t) = b_x(t) b_y(t) b_z(t)$.
The ${\bf D}$ vector can be written as
\begin{equation}
\label{eqn:D-time}
D_j({\bf R}(t), t) = \frac{1}{\sqrt{A(t)}}
\phi_j({\bf R}(0),t) \exp{\left[i S ({\bf R}(t), t)\right]},
\nonumber
\end{equation}
which upon substitution into Eq. (\ref{eqn:GPunitary})
and Eq. (\ref{eqn:staticGP}) at $t=0$ leads to
\begin{equation}
\label{eqn:phase}
S({\bf R}(t), t) = S_0(t) + M \sum_{k} \frac{R_k^2(t) }{b_k(t)} \frac{d b_k(t)}{dt}.
\end{equation}
For a cigar-shaped trapping potential with axial symmetry
($\epsilon \equiv \omega_z(0) / \omega_\perp(0) \ll 1$
and $\omega_\perp \equiv \omega_x = \omega_y$), $D_j({\bf R},t)$
becomes
\begin{equation}
\label{eqn:Dlimit}
D_j({\bf R},t) \approx \frac{\exp[iS({\bf R},t)]}{\sqrt{1+ \lambda^2}}
D_j(\overline{\bf R},0),
\end{equation}
where $\overline{R}_k = R_k/b_k(t)$ are scaled coordinates, and
$\lambda \equiv \omega_\perp(0) t$ is the dimensionless time. The
result for $s$-wave is formaly identical to that of
Eq.~(\ref{eqn:Dlimit}) with the substitution $D_j \to \Psi$, where
$\Psi$ represents the scalar order parameter.

The matter-wave interference of two spatially separated condensates (without tunneling)
is described by
\begin{equation}
\label{eqn:coherentstate}
\Phi_{\rm tot}({\bf R},t) = \Phi_{\rm
L}({\bf R},t) + \Phi_{\rm R} ({\bf R},t),
\end{equation}
where $\Phi_{\rm P}\propto i \sum_j D_{j,{\rm P}} \sigma_j
\sigma_y$ denotes the pair wavefunction of Fermi condensates in
the left [${\rm P}={\rm L}(+)$] or right [R$(-)$] traps. Here, a
coordinate system is chosen such that the trap centers lie at
$(-W/2, 0,0)$ and $(W/2,0,0)$, where $W$ is the distance between
the traps. We consider the case of two identical cigar shaped
Fermi condensates, with the weakest confinement along the ${\bf
z}$ axis, just like the experiment in Bose systems~\cite{shin-04}.
In this case, the time evolution of the two independent Fermi
condensates is described by
\begin{equation}
\label{eqn:Dlt}
D_{j,{\rm P}} ({\bf R},t) = \frac{\exp[iS({\bf R}\pm W\hat{\bf x}/2,t)]}{\sqrt{1+ \lambda^2}}
D_{j} (\overline{{\bf R}\pm W\hat{\bf x}/2},0),
\nonumber
\end{equation}
Thus, for any single run of the experiment, the particle density
$n({\bf R},t)\equiv \vert \Phi_{\rm tot}({\bf R},t) \vert^2$ is
\begin{eqnarray}
\label{eqn:n}
n({\bf R}, t) &\propto& \vert {\bf D}_{\rm L}({\bf R},t) \vert^2
+ \vert {\bf D}_{\rm R}({\bf R},t) \vert^2
\nonumber \\
&& \hspace{-2cm}
+ 2 {\rm Re}\left[
\frac{{\bf D}^\dagger (\overline{{\bf R}+W\hat{\bf x}/2},0)
\cdot {\bf D}(\overline{{\bf R}-W\hat{\bf x}/2},0)}
{\sqrt{A(\lambda) A(\lambda)}}
e^{i\chi} \right],
\end{eqnarray}
where the phase $\chi({\bf R},t) = S({\bf R}+W\hat{\bf x}/2,t )-
S({\bf R}-W\hat{\bf x}/2,t )+ \chi_0$, and $\chi_0$ is the initial
relative phase of the two clouds. The result for $s$-wave is
formaly identical to that of Eq.~(\ref{eqn:n}) with the
substitution $D_j \to \Psi$.

In the case where both Fermi condensates are in unitary states,
${\bf D}$ is essentially a real vector with an overall phase, and
$n({\bf R},t)$ shows an angular dependence controlled by the dot
product in Eq.~(\ref{eqn:n}). When the two order parameters are
parallel, this term is maximal and the interference pattern is
most visible (Fig. 2). However if the ${\bf D}$ vectors are
perpendicular, fringes are absent at all times (Fig. 3).
Therefore, in the unitary case the existence and intensity of
interference fringes are very sensitive to the relative
orientation of the vector order parameters.
\begin{figure}
\begin{center}
\includegraphics[width=3.5cm]{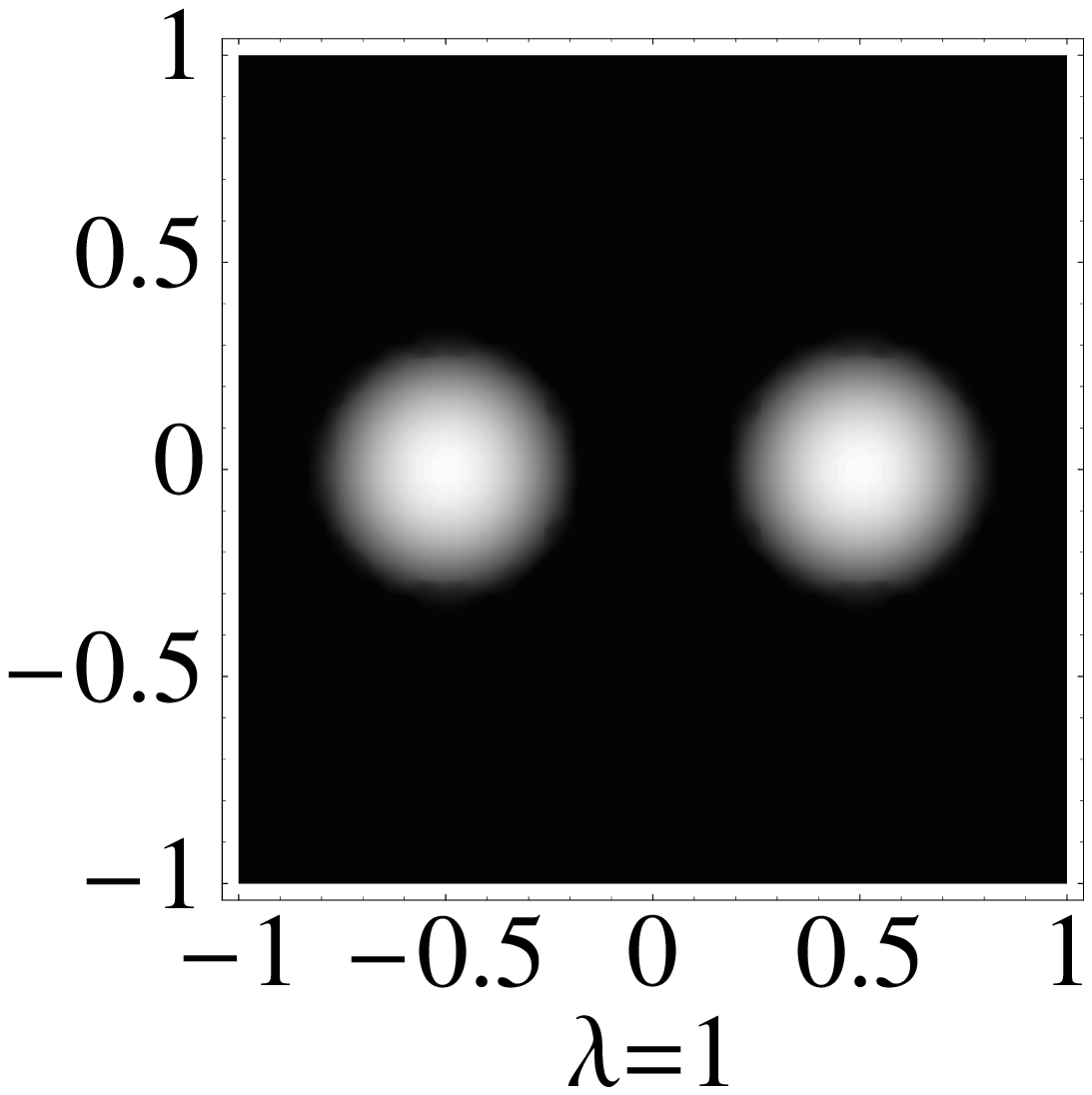}
\hspace{0mm}
\includegraphics[width=3.5cm]{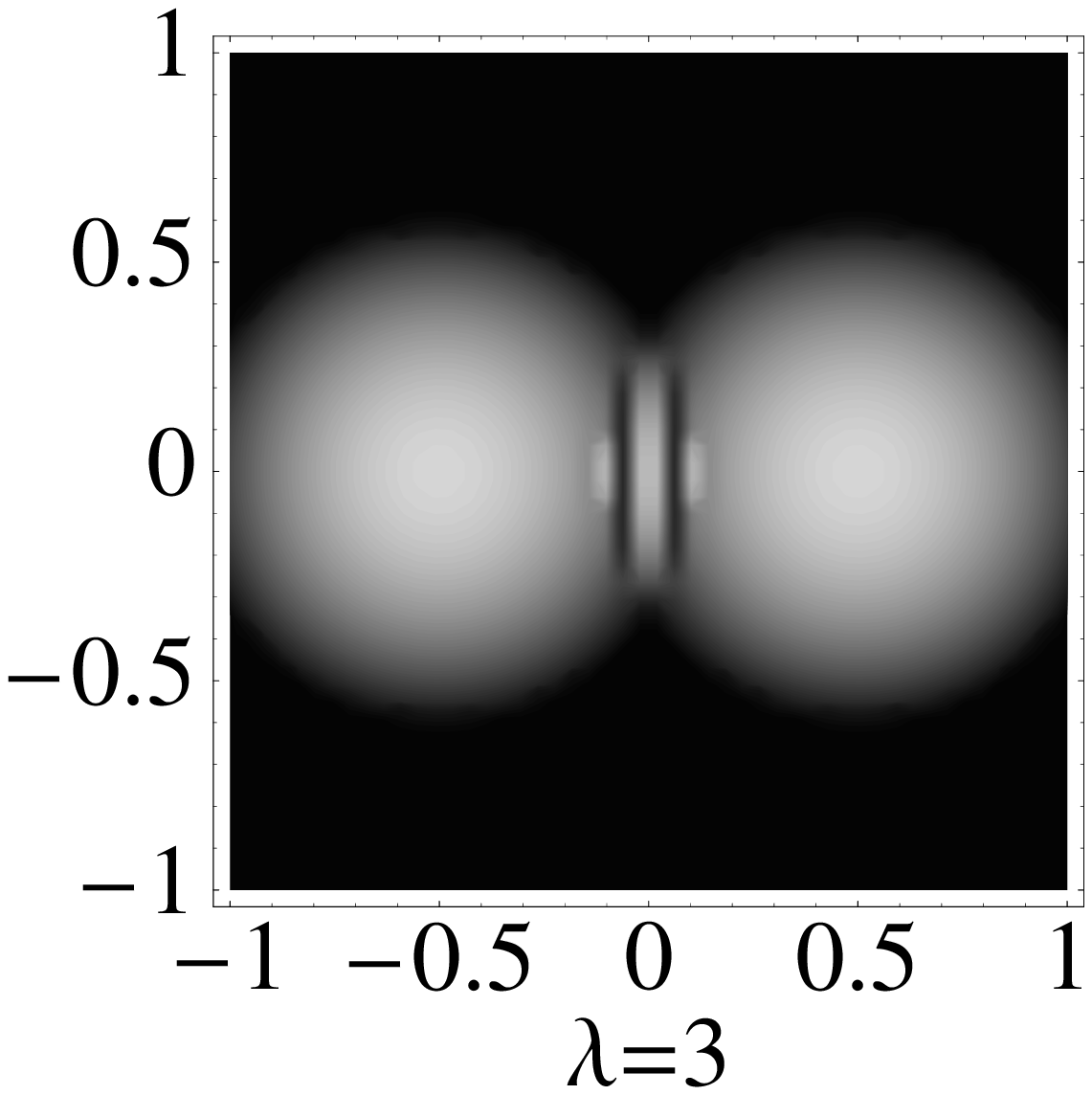}
\vskip 2mm
\includegraphics[width=3.5cm]{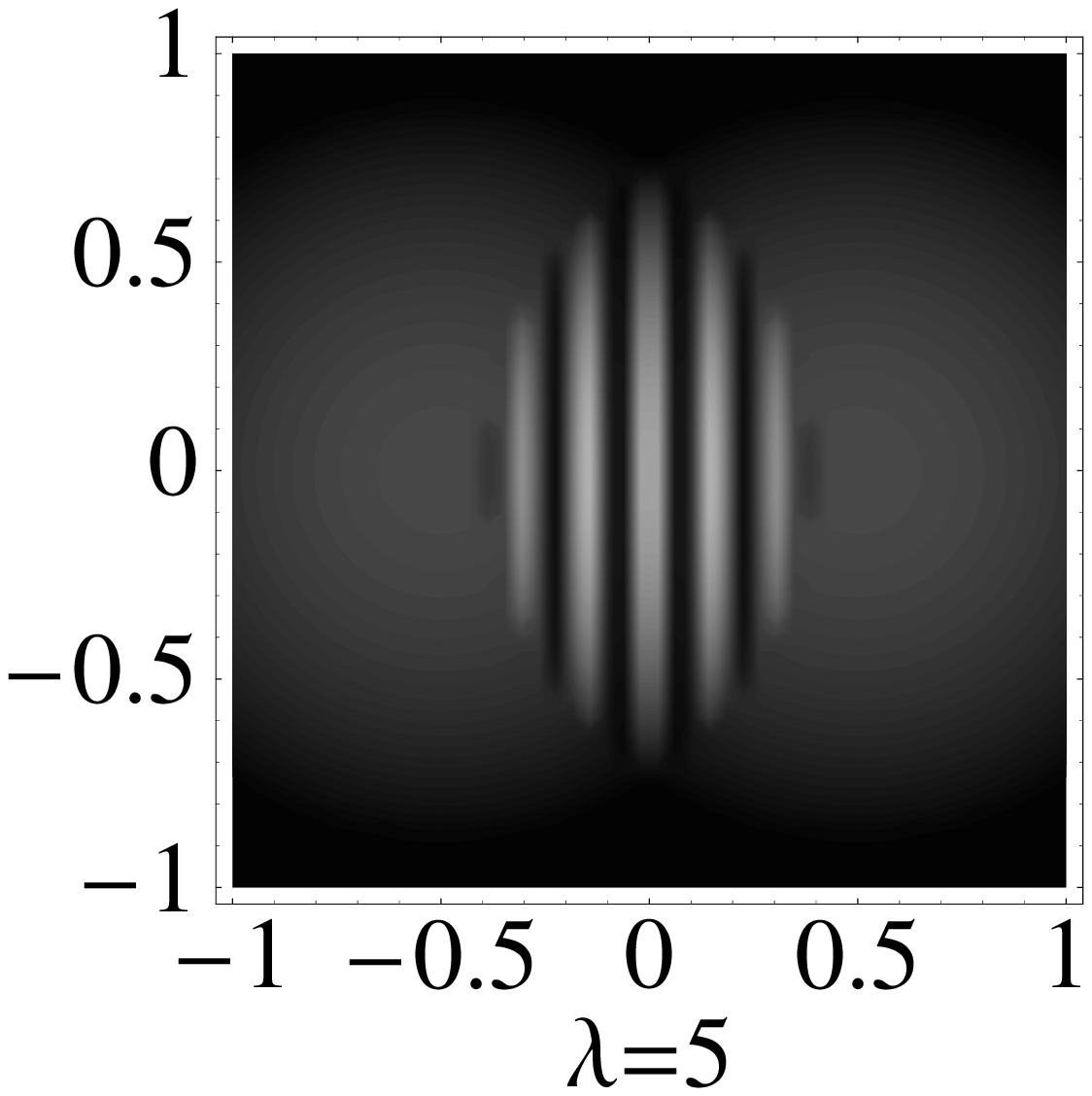}
\hspace{0mm}
\includegraphics[width=3.5cm]{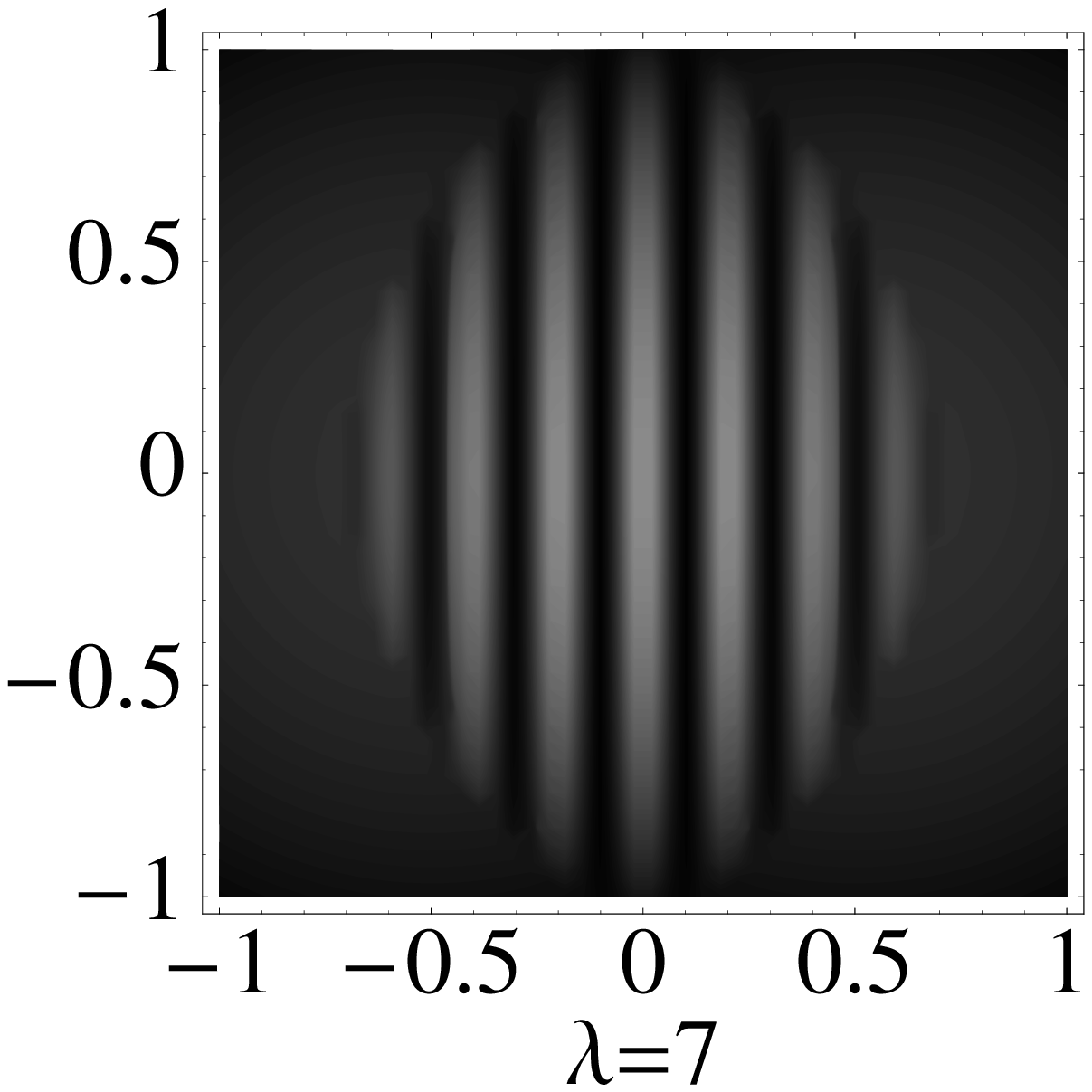}
\end{center}
\caption{Interference pattern versus dimensionless time $\lambda$ for unitary $p$-wave Fermi condensates
in the BEC limit involving two hyperfine states (cross section view). Each condensate is initially trapped
in a cigar-shaped potential with ${\bf D}_{\rm L}^{\dagger} \parallel {\bf D}_{\rm R}$.
The patterns are similar for atomic scalar bosons, and for $s$-wave and single hyperfine state $p$-wave
fermions.}
\label{fig:inter-isopara}
\end{figure}
\begin{figure}
\begin{center}
\includegraphics[width=3.5cm]{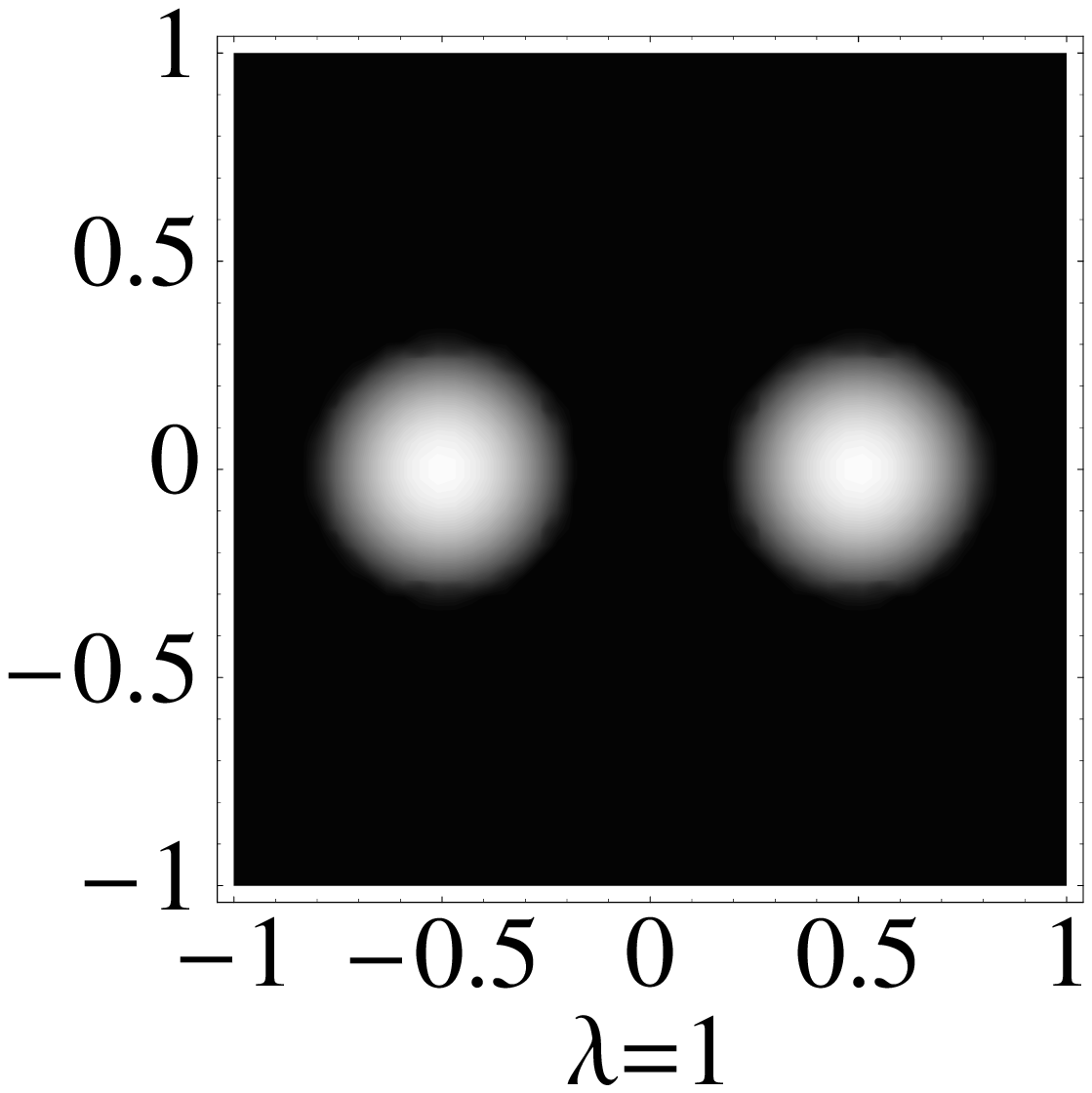}
\hspace{0mm}
\includegraphics[width=3.5cm]{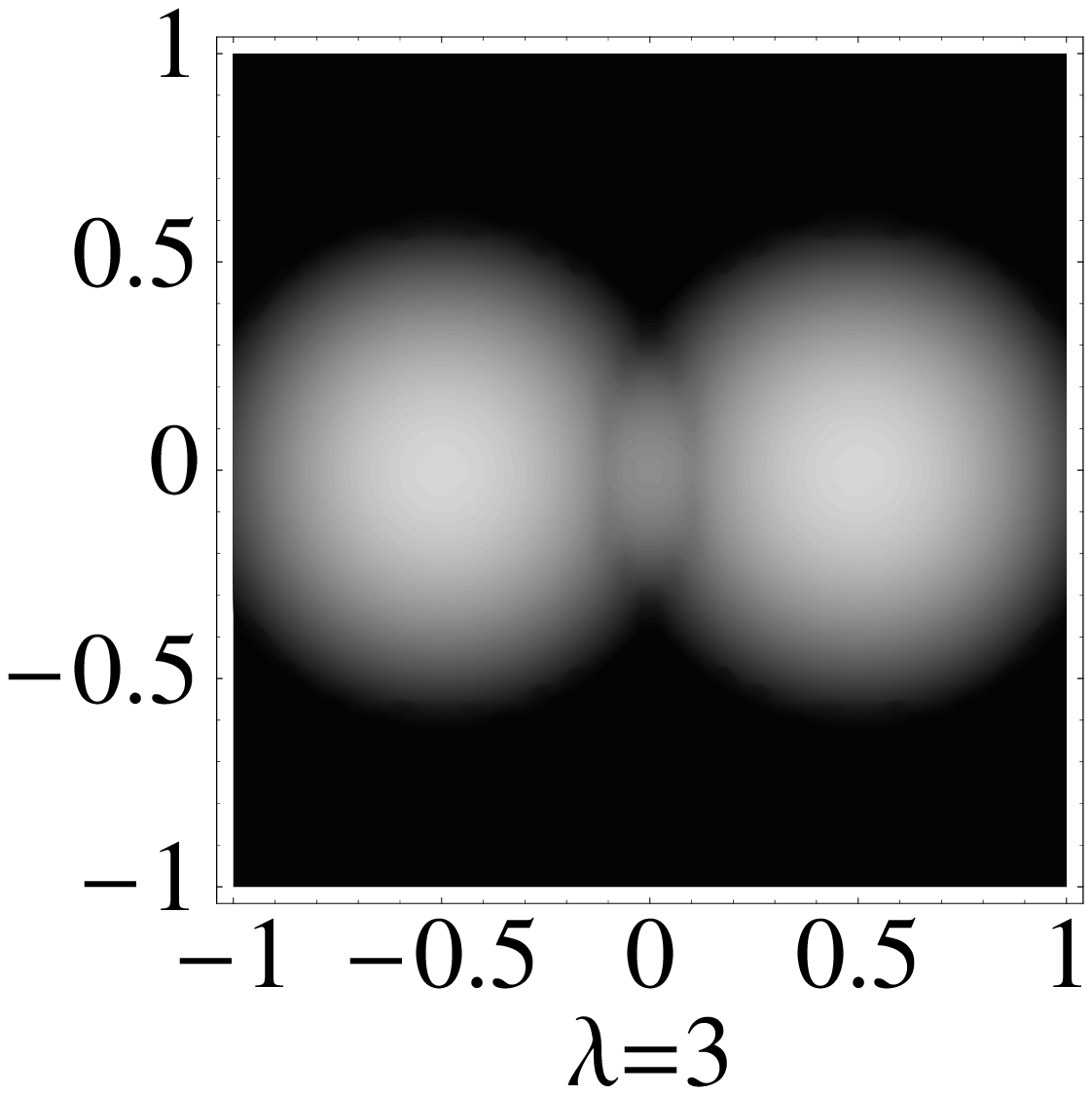}
\vskip 2mm
\includegraphics[width=3.5cm]{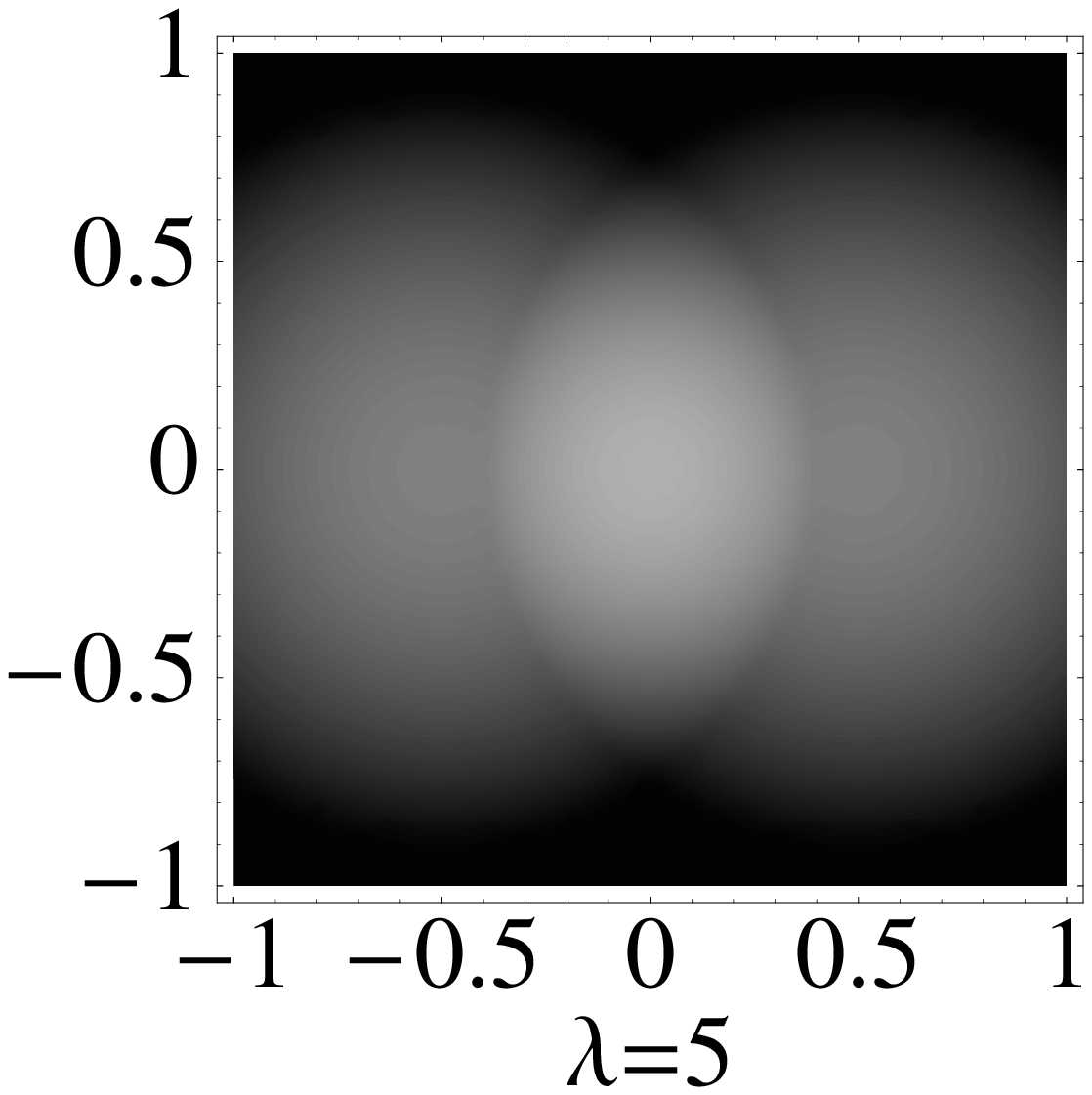}
\hspace{0mm}
\includegraphics[width=3.5cm]{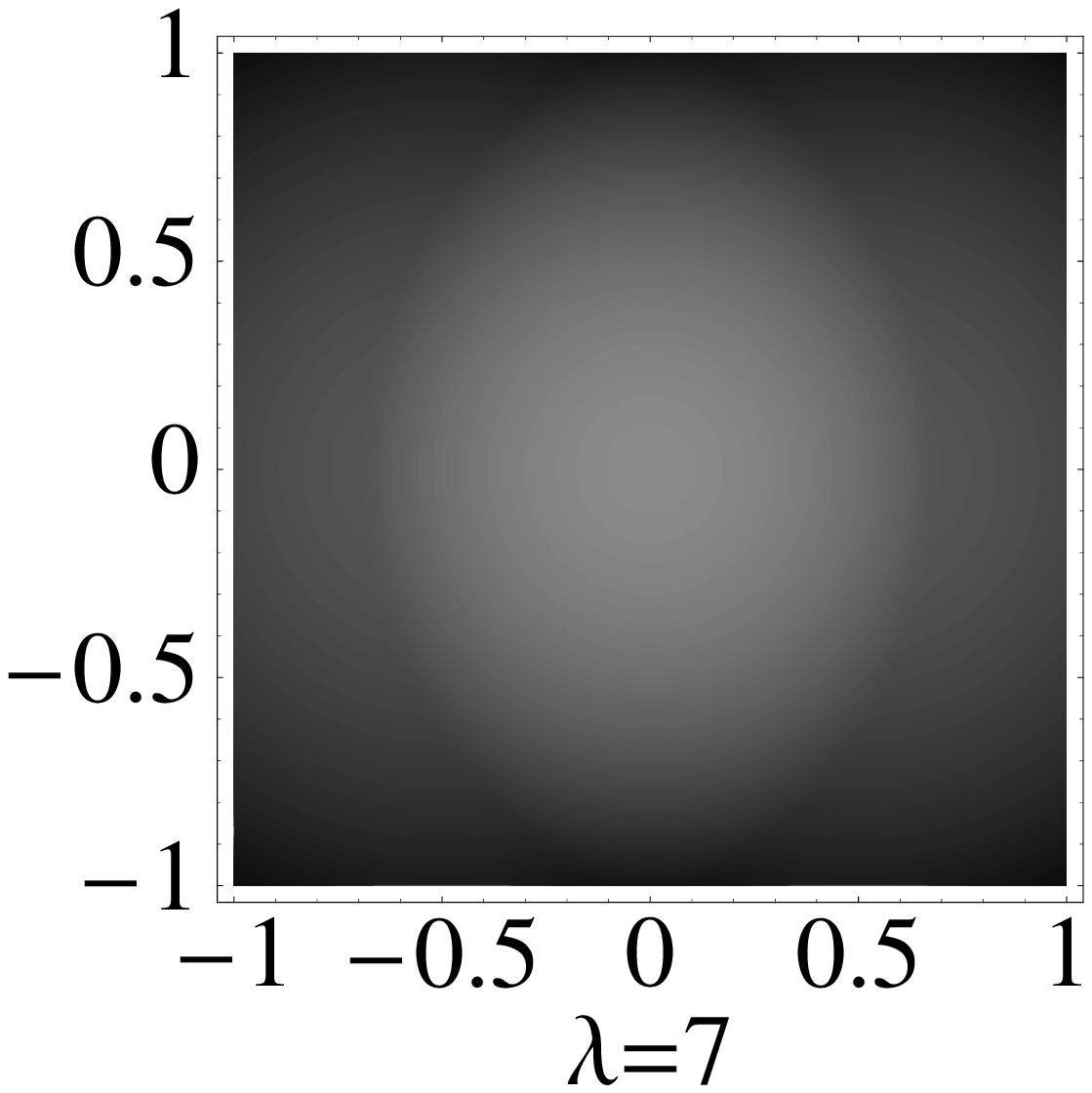}
\end{center}
\caption{Same plots as in Fig. 2, but with ${\bf D}_{\rm L}^{\dagger} \perp {\bf D}_{\rm R}$.}
\label{fig:inter-isoperp}
\end{figure}

When only one hyperfine state is occupied,
the ${\bf D}$ vector has a fixed form $A(1,i,0)$.
Without this pseudospin space rotational degree of freedom, the angular dependence of the
interference pattern disappears and fringes are present in all experimental realizations.
This result is similar to the $s$-wave case where the order parameter is a complex scalar.

Before concluding, we would like to mention two important points.
First, away from the BEC limit the kinetic energy term in Eq. (\ref{eqn:GP}) is
$$
\label{eqn:massaniso}
-\sum_{i,j,m_\ell} \frac{\nabla_i \nabla_j}{2M_{m_\ell}^{i,j}} D_{j,m_\ell},
$$
thus acquiring a mass anisotropy $M_{m_\ell}^{i,j} =
M/c_{m_\ell}^{i,j}$, which is directly related to the anisotropy
of the Ginzburg--Landau coherence length $\xi_{ij}$. Here, we use
the basis of spherical harmonics $Y_{1,m_\ell}$ and define $D_{j,
m_\ell}$ by $d_j({\bf r}, {\bf r}^\prime) = \sum_{m_\ell}
D_{j,m_\ell}({\bf R}) \eta_{0,m_\ell}(x) Y_{1,m_\ell}(\hat{\bf
x})$. For a weak trapping potential, the coefficient
$c_{m_\ell}^{i,j}$ becomes
\begin{eqnarray}
\label{eqn:cij}
c_{m_\ell}^{i,j} &=& \sum_{{\bf k},m_\ell^\prime}
\Big\{ \left[\frac{X({\bf k})}{4 E^2({\bf k})} - \frac{\beta Y({\bf k})}{16 E({\bf k})}\right]
\delta_{m_\ell, m_\ell^\prime} \delta_{i,j}
\nonumber \\
&+&
\kappa_{m_\ell,m_\ell^\prime}^{i,j}
\frac{\beta^2 {\bf k}^2 X({\bf k}) Y ({\bf k})}{32 m E({\bf k})}
\Big\} \phi^2(k),
\end{eqnarray}
where $E({\bf k}) = \xi_{1,{\bf k}} + \xi_{2,{\bf k}}$,
$\xi_{\alpha,{\bf k}} = {\bf k}^2/2m - \mu_\alpha$,
$X({\bf k}) = \tanh(\beta \xi_{1,{\bf k}}/2) + \tanh(\beta\xi_{2,{\bf k}}/2)$,
and $Y({\bf k}) = {\rm sech}^2(\beta\xi_{1,{\bf k}}/2)
+ {\rm sech}^2(\beta\xi_{2,{\bf k}}/2)$.
The symmetry function $\phi(k)$ is defined by
$V({\bf k}, {\bf k}') = \int d{\bf x} V({\bf x}) \exp [i({\bf k}-{\bf k}')\cdot {\bf x}]
= V \phi(k) \phi(k') Y_{1,m_\ell}(\hat{\bf k}) Y_{1,m_\ell^\prime}^*(\hat{\bf k}')$,
and the angular average
\begin{equation}
\kappa_{m_{\ell}, m_\ell^\prime}^{i,j} =
\int d\hat{\bf k} \hat{k}_i \hat{k}_j
Y_{1,m_\ell}(\hat{\bf k}) Y_{1,m_{\ell}^\prime}^*(\hat{\bf k}).
\nonumber
\end{equation}
When the order parameter is characterized by one $Y_{1,m_\ell}$
with $m_{\ell}$ fixed, $\kappa$ is diagonal in both $m_\ell$ and
$i$, hence $c_{m_\ell}^{i,j} = c_{j,m_\ell} \delta_{i,j}$. This
mass anisotropy reflects the higher angular momentum ($p$-wave)
nature of the order parameter for paired fermions, and it is
completely absent in $s$-wave Fermi and atomic Bose condensates.

This effective mass anisotropy has a non-trivial influence on
the time evolution of condensates after release from the trap.
Making the scale transformation
$R_j^\prime =  R_j \sqrt{M_{j,m_\ell}}$,
$\omega_j^\prime =\omega_j/\sqrt{M_{j,m_\ell}}$,
we conclude that the cloud expansion predominantly occurs in the strongest confined
direction in the scaled space. Thus, the
time evolution of a Fermi condensate initially trapped in an axially
symmetric potential with sizes $L_x = L_y < L_z$ leads to
an anisotropic cross section of the cloud at any time after release
when $M_x \neq M_y$ ($\xi_x \neq \xi_y$). In Fig. 4, we show the cloud anisotropy ratio
as a function of the effective mass anisotropy ratio.
\begin{figure}
\begin{center}
\psfrag{r}{$r_L$}
\psfrag{rM}{$r_M$}
\includegraphics[width=5.4cm]{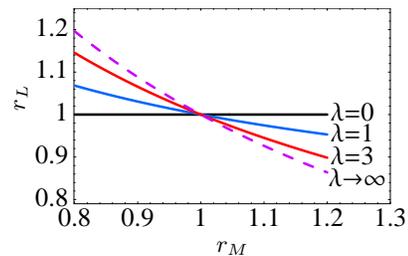}
\end{center}
\caption{Cloud anisotropy ratio $r_L = L_x/L_y$ as a function of
effective mass anisotropy ratio $r_M = M_x/M_y$ at time $\lambda$ (Solid lines).
Dashed line indicate the saturated behavior at $\lambda \to \infty$.}
\label{fig:ratio}
\end{figure}

Second, in the discussion above we assume a symmetric interaction in pseudospin space,
i.e., $V_{1111}$, $V_{1212}$ and $V_{2222}$ are identical.
However, experimental results for $p$-wave Feshbach resonances show a
a small but finite separation in different channels~\cite{schunck-05}.
Different interactions in pseudospin space
do not affect dramatically our derivation or results.
When the different interactions are absorbed into an effective ${\bf D}$ vector,
an equivalent procedure leads to equations similar to Eqs.(5) and (6).

In conclusion, we considered a Fermi condensate consisting of two
hyperfine states with $s$-wave and $p$-wave interactions,
and derived equation of motion for $p$-wave case in a vector boson representation
near the BEC limit. We found that the quantum interference of two $p$-wave Fermi condensates
has a polarization effect due to the vector nature of the order parameter.
This effect is absent in strongly coupled (BEC) $s$-wave Fermi superfluids,
as well as in scalar atomic Bose systems.
Furthermore, we observed that different orbital symmetries of the
vector order parameter produce anisotropic effective masses (coherence lengths).
For a cigar-shaped cloud with an isotropic cross section,
the cloud expansion becomes anisotropic at any time after release
from the trap when the effective masses are anisotropic.
Thus, the orbital symmetry of the order parameter
for $p$-wave condensates can be probed via single cloud expansions.


\begin{references}

\bibitem{shin-04}
Y. Shin {\it et. al.},
%M. Saba, T. A. Pasquini, W. Ketterle, E. D. Pritchard, and A. E. Leanhardt,
Phys. Rev. Lett. {\bf 92}, 050405 (2004).

\bibitem{schumm-05}
T. Schumm {\it et. al.},
%S. Hofferberth, L. M. Andersson, S. Wildermuth, S. Groth,
%I. Bar-Joseph, J. Schmiedmayer, and P. Kruger,
Nature Physics {\bf 1}, 57 (2005).

\bibitem{bloch-05}
S. Folling {\it et. al.},
Nature {\bf 434}, 481 (2005).

\bibitem{hulet-03}
K. E. Strecker, G. B. Partridge, and R. G. Hulet, Phys. Rev. Lett. {\bf 91}, 080406 (2003).

\bibitem{greiner-03}
M. Greiner, C. A. Regal, and D. S. Jin, Nature (London) {\bf 426},
537 (2003).

\bibitem{zwierlein-03}
M. W. Zwierlein {\it et. al.},
%C. A. Stan, C. H. Schunck, S. M. F. Raupach,
%S. Gupta, Z. Hadzibabic, and W. Ketterle,
Phys. Rev. Lett. {\bf 91}, 250401 (2003).

\bibitem{bartenstein-04}
M. Bartenstein {\it et. al.},
%A. Altmeyer, S. Riedl, S. Jochim, C. Chin,
%J. Hecker Denschlag, and R. Grimm,
Phys. Rev. Lett. {\bf 92}, 120401 (2004).

\bibitem{bourdel-04}
T. Bourdel {\it et. al.},
%L. Khaykovich, J. Cubizolles, J. Zhang,
%F. Chevy, M. Teichmann, L. Tarruell, S. J. J. M. F. Kokkelmans, and C. Salomon,
Phys. Rev. Lett. {\bf 93}, 050401 (2004).

\bibitem{thomas-04}
J. Kinast {\it et. al.},
%S. L. Hemmer, M. E. Gehm, A. Turlapov and J. E. Thomas,
Phys. Rev. Lett {\bf 92}, 150402 (2004).

\bibitem{sademelo-93}
C. A. R. S{\'a} de Melo, M. Randeria, and J. R. Engelbrecht,
Phys. Rev. Lett. {\bf 71}, 3202 (1993).

\bibitem{iskin-06}
M. Iskin and C. A. R. S{\'a} de Melo, Phys. Rev. Lett. {\bf 96}, 040402 (2006).

\bibitem{regal-03b}
C. A. Regal, C. Ticknor, J. L. Bohn, and D. S. Jin,
Phys. Rev. Lett. {\bf 90}, 053201 (2003).

\bibitem{zhang-04}
J. Zhang {\it et. al.},
%E. G. M. van Kempen, T. Bourdel, L. Khaykovich, J. Cubizolles,
%F. Chevy, M. Teichmann, L. Tarruell, S. J. J. M. F. Kokkelmans, and C. Salomon,
Phys. Rev. A {\bf 70}, 030702(R) (2004).

\bibitem{schunck-05}
C. H. Schunck {\it et. al.},
%M. W. Zwierlein, C. A. Stan, S. M. F. Raupach,
%W. Ketterle, A. Simoni, E. Tiesinga, C. J. Williams, and P. S. Julienne,
Phys. Rev. A {\bf 71}, 045601 (2005).

\bibitem{gunter-05}
K. Gunter {\it et. al.},
%T. Stoferle, H. Moritz, M. Kohl, and T. Esslinger,
Phys. Rev. Lett. {\bf 95}, 230401 (2005).

\bibitem{tokatly-04}
I. V. Tokatly, Phys. Rev. A {\bf 70}, 043601 (2004).

\bibitem{kagan-96}
Yu. Kagan, E. L. Surkov, and G. V. Shlyapnikov,
Phys. Rev. A {\bf 54}, R1753 (1996).

\bibitem{castin-96}
Y. Castin and R. Dum, Phys. Rev. Lett. {\bf 77}, 5315 (1996).

\end{references}
\end{document}